\documentclass[sigconf]{acmart}

\usepackage{amsmath,amsfonts} %amssymb,
\usepackage{graphicx}
\usepackage{textcomp}
\usepackage{xcolor}
\usepackage{balance}
\usepackage{color, colortbl}
\definecolor{Gray}{gray}{0.9}

\usepackage[noend]{algpseudocode}
\usepackage{algorithm,algorithmicx}
\usepackage{xfrac}

\algdef{SE}[SUBALG]{Indent}{EndIndent}{}{\algorithmicend\ }
\algtext*{Indent}
\algtext*{EndIndent}

\algdef{SE}[DOWHILE]{Do}{doWhile}{\algorithmicdo}[1]{\algorithmicwhile\ #1}%

\def\BibTeX{{\rm B\kern-.05em{\sc i\kern-.025em b}\kern-.08em
    T\kern-.1667em\lower.7ex\hbox{E}\kern-.125emX}}

\newcommand{\mA}{\mathbf{A}} 
\newcommand{\mC}{\mathbf{C}} 
\newcommand{\mL}{\mathbf{L}} 
 
\newcommand{\mR}{\mathbf{R}}
 
\newcommand{\mP}{\mathbf{P}} 

\newcommand{\mS}{\mathbf{S}}
\newcommand{\md}{\mathbf{d}}
\newcommand{\mpre}{\mathbf{pre}}
\newcommand{\mpost}{\mathbf{post}}

\newcommand{\mb}{\mathbf{b}}

\newcommand{\mv}{\mathbf{v}} 
\newcommand{\mpar}{\mathbf{p}} 
\newcommand{\transpose} {^{\mbox{\scriptsize \sf T}}}

\newcommand{\kmer}[1][]{\textit{k}-mer#1}

\copyrightyear{2022}
\acmYear{2022}
\setcopyright{rightsretained}
\acmConference[ICPP '22]{51st International Conference on Parallel
Processing}{August 29-September 1, 2022}{Bordeaux, France}
\acmBooktitle{51st International Conference on Parallel Processing (ICPP
'22), August 29-September 1, 2022, Bordeaux,
France}\acmDOI{10.1145/3545008.3545050}
\acmISBN{978-1-4503-9733-9/22/08}

\begin{document}

\title{Distributed-Memory Parallel Contig Generation for \emph{De Novo} Long-Read Genome Assembly}

\author{Giulia Guidi}
\authornote{Both authors contributed equally to this research.}
\affiliation{%
    \institution{%\fontsize{9}{11}\selectfont 
University of California, Berkeley}
    \institution{%\fontsize{9}{11}\selectfont 
Lawrence Berkeley National Laboratory}
    \city{Berkeley}
    \state{California}
    \country{USA}
}

\author{Gabriel Raulet}
\authornotemark[1]
\affiliation{%
    \institution{%\fontsize{9}{11}\selectfont 
Lawrence Berkeley National Laboratory}
    % \institution{and Google, Inc.}
    \city{Berkeley}
    \state{California}
    \country{USA}
}

\author{Daniel Rokhsar}
\affiliation{%
    \institution{%\fontsize{9}{11}\selectfont 
University of California, Berkeley}
    \institution{%\fontsize{9}{11}\selectfont 
DOE Joint Genome Institute}
    \city{Berkeley}
    \state{California}
    \country{USA}
}

\author{Leonid Oliker}
\affiliation{%
    \institution{%\fontsize{9}{11}\selectfont 
Lawrence Berkeley National Laboratory}
    % \institution{and Google, Inc.}
    \city{Berkeley}
    \state{California}
    \country{USA}
}

\author{Katherine Yelick}
\affiliation{%
    \institution{%\fontsize{9}{11}\selectfont 
University of California, Berkeley}
    \institution{%\fontsize{9}{11}\selectfont 
Lawrence Berkeley National Laboratory}
    \city{Berkeley}
    \state{California}
    \country{USA}
}

\author{Ayd{\i}n Bulu\c{c}}
\affiliation{%
    \institution{%\fontsize{9}{11}\selectfont 
University of California, Berkeley}
    \institution{%\fontsize{9}{11}\selectfont 
Lawrence Berkeley National Laboratory}
    \city{Berkeley}
    \state{California}
    \country{USA}
}

\begin{abstract}
\emph{De novo} genome assembly, i.e., rebuilding the sequence of an unknown genome from redundant and erroneous short sequences, is a key but computationally intensive step in many genomics pipelines.
The exponential growth of genomic data is increasing the computational demand and requires scalable, high-performance approaches.

In this work, we present a novel distributed memory algorithm that, from a string graph representation of the genome and using sparse matrices, generates the contig set, i.e., overlapping sequences that form a map representing a region of a chromosome.

Using matrix abstraction, we mask branches in the string graph, and compute the connected component to group genomic sequences that belong to the same linear chain (i.e., contig).
Then, we perform multiway number partitioning to minimize the load imbalance in local assembly, i.e., concatenation of sequences from a given contig.
Based on the assignment obtained by partitioning, we compute the induce subgraph function to redistribute sequences between processes, resulting in a set of local sparse matrices.
Finally, we traverse each matrix using depth-first search to concatenate sequences.

Our algorithm shows good scaling with parallel efficiency up to 80\% on 128 nodes, resulting in uniform genome coverage and showing promising results in terms of assembly quality. 
Our contig generation algorithm localizes the assembly process to significantly reduce the amount of computation spent on this step.
Our work is a step forward for efficient \emph{de novo} long read assembly of large genomes in a distributed memory.

\end{abstract}

\maketitle

\section{Introduction}

Genomics is facing exponential growth in data coming from improved and cheaper instrumentation that is overwhelming conventional analytical infrastructures. 
It is especially true for \emph{de novo} genome assembly~\cite{zhang2011practical}, that is decoding the sequence of an unknown genome from redundant and erroneous short sequences. 
A key but computationally intensive step in many genomics pipelines.

On the one hand, large-scale parallel systems promise to overcome the computational and memory limitation of conventional shared-memory computing. 
On the other hand, programming such large-scale systems is challenging, especially when computation is highly irregular, as is often the case in genomics.

The advent of long--read sequencing technologies~\cite{eid2009real, goodwin2015oxford}, which yield sequences with an average length of more than 10,000 base pairs (bp), improved our ability to assemble complex genomic repeats and obtain more accurate assemblies that were not possible with short--read technologies~\cite{phillippy2008genome, nagarajan2009parametric}.
In the past, longer sequences were associated with higher error rates~\cite{hon2020highly}, resulting in significantly higher algorithm complexity and computational costs.
Recent advances in sequencing technologies have lowered error rates, reducing the need for error correction and polishing phases, but computational costs remain high as we generate more data and reconstruct higher quality assemblies.

It is common practice to assemble long-read data according to the  Overlap--Layout--Consensus (OLC) paradigm~\cite{berlin2015assembling}.
The first step (O) is to identify overlaps between reads to create an \emph{overlap} graph. 
The second step (L) simplifies the overlap graph by removing redundant information and transforming it into an \emph{string} graph.
The third step (C) takes the string graph and processes it to obtain a first draft of the unknown genome, returning a contig set as output. 
A \emph{contig} is a group of overlapping DNA sequences that together form a region of a chromosome.
A common approach to contig generation takes the string graph as input, branching nodes in this graph are masked, and the set of linear, unbranched paths in the graph is extracted to form the contig set.
This step can be followed by further polishing phases to merge and correct the contig set and to separate haplotypes.

In this paper, we build on previous work in the literature, diBELLA 2D~\cite{guidi2021parallel, guidi2021bella}, and present a novel distributed-memory algorithm that generates the contig set starting from a string graph representation of the genome and using a sparse matrix abstraction.
The proposed approaches are implemented in a software package called ELBA, which also includes the overlap and layout phases of diBELLA 2D, which can scale to hundreds of nodes. 
% \Giulia{
Each stage of the computation in ELBA is implemented for distributed memory.
% }
% (see Section~\ref{sec:back}) 

The algorithm first determines the contig set starting from a string graph by masking out the branches and extracting the non-branched paths by computing connected components over the graph (i.e., a sparse matrix).
Using a greedy multiway number partitioning algorithm, ELBA then determines how to redistribute sequences (i.e., rows and columns of the sparse matrix) between processes so that sequences belonging to the same contig are stored on the same processor.
Once the sequences-to-processor assignment is computed, ELBA uses this information to implement the induced subgraph function that redistributes the sequences using sparse matrix abstraction so that each processor has a local sparse matrix representing the non-branching contig(s) assigned to the processor.
ELBA then computes a local assembly step on each processor independently and in parallel. In doing so, it uses depth-first search to concatenate the sequences belonging to a particular contig and return the contig set.
ELBA is the first distributed-memory implementation of the OLC assembly paradigm.

Our algorithm shows good scaling with parallel efficiency up to 80\% on 128 nodes on a representative dataset. 
Our approach leads to uniform coverage of the genome and shows promising results in terms of assembly quality. 
Our contig generation algorithm is fast and efficient because it localizes the graph traversal problem, i.e., the practical concatenation of sequences, so that most of the computation can be performed locally without requiring fine-grained communication.
Compared to related work, this approach significantly reduces the fraction of the runtime required by the entire assembly pipeline for contig generation.
Our implementation is publicly available at \url{https://github.com/PASSIONLab/ELBA}.

\vspace{-.5em}
\section{Background}\label{sec:back}

% \paragraph{Genome and Genome Assembly}

% In humans, for example, each cell normally contains 23 pairs of chromosomes, for a total of 46.
A genome consists of one or more linear chains of DNA molecules, i.e., nucleotides or bases, represented by the alphabet $\Sigma = \{\rm A, C, G,$ $T\}$.
Each linear chain has a direction and is paired with a second chain called its \emph{reverse complement}, which has the opposite direction.
Together they are called \emph{strands} and wind around each other to form a double helix; this double helix is organized in three-dimensional space and forms a chromosome. 
Different species may have different numbers of chromosomes.
On the opposite strands, A always pairs with T and C pairs with G. 
Given a string $v = \rm ATTCG$, its reserve complement is $v' = \rm CGAAT$.
% If $v = \rm ATTCG$, 
% The \emph{canonical form} of a DNA sequence $v$ is the lexicographically smaller of $v$ and its reverse complement $v'$. 
% In our example, $v = \rm ATTCG$ is the canonical form.

% OLC is the most widely used assembly paradigm for long read data~\cite{berlin2015assembling,li2016minimap}.
% Its first step consists of identifying overlaps between input sequences.
% The idea behind the search for overlaps is that two sequences that overlap may originate from adjacent positions on the genome.
% However, the assembly process is more complex than it might seem at first glance.
% This is because we cannot be sure that two overlapping sequences actually originate from adjacent positions on the genome due to the repetitiveness of the genome.

The goal of \emph{de novo} long-read genome assembly is to reconstruct an unknown genome sequence, possibly consisting of multiple chromosomes, starting from a collection of redundant, erroneous, and significantly shorter sequences called \emph{reads}.
For long-read sequencing data, this is commonly done following the OLC assembly paradigm introduced in the previous section.
In this paper, we take the overlap (O) and layout (L) phases as implemented in diBELLA 2D~\cite{guidi2021parallel, guidi2021bella} and extend them to include the third (C) phase, that is contig generation.
diBELLA 2D uses a $k$-mer--based approach and relies on parallel sparse matrix multiplication for overlap detection and transitive reduction to obtain a string graph (i.e., the input to the contig generation algorithm described in Section~\ref{sec:algortihm}).

diBELLA 2D sees the overlap detection step as a sparse matrix multiplication between a $\lvert\textit{sequences}\rvert$-by-$\lvert\textit\kmer{s}\rvert$ matrix $\mA$ and its transpose $\mA\transpose$.
The resulting matrix $\mC = \mA\mA\transpose$ is a sparse $\lvert\textit{sequences}\rvert$-by-$\lvert\textit{sequences}\rvert$ that stores information about which sequences share at least one $k$-mer.
diBELLA 2D then computes an element-wise pairwise alignment step for each non-zero in the $\mC$ matrix, removing non-zeros that are below a certain alignment threshold.
The result of this operation is the result matrix $\mR$, which in turn is the input for the transitive reduction phase, whose goal is to remove redundancies from the overlap graph and return a string graph in the form of a sparse matrix $\mS$.

A string graph (or matrix) is a graph $G = (V, E)$, where $V$ is the set of sequences and $E$ is the set of overlap \emph{suffixes} between any two vertices.
% There is an edge if and only if the respective reads overlap and the weight of this edge is the length of the suffix.
A string graph contains neither redundant edges nor redundant vertices. 
A redundant vertex is a read that is completely contained (i.e., aligned) within another read and whose edges therefore contain information that the containing read already has.
A redundant edge or transitive edge is an edge that carries less or the same information as a parallel path in the graph and therefore can be removed without loss of information during transitive reduction.

In diBELLA 2D, so also in ELBA, the string graph is encoded as a bidirected graph, so we can encode information about both DNA strands.
A {\it bidirected graph}~\cite{edmonds2003matching} is one in which each edge at each end has an independent orientation (or arrow). 
There are three types of bidirected edges: (a) those with arrows at both ends pointing outward, (b) those with both arrows pointing inward, and (c) those with both arrows pointing in the same direction, one away from its vertex and the other toward its vertex.

% For example, for the sequences $v_1 = \rm TACGA$ and $v_2 = \rm ACGACC$, their overlap suffix or \emph{overhang} is the portion of $v_2$ that exceeds the overlap between $v_1$ and $v_{2}$, i.e. $e_{12} = \rm CC$. 
% Given $G = (V, E)$, where $V = \{v_1, v_2, v_3\}$ and $E = \{e_{12}, e_{13}, e_{23}\}$, we can walk (a) $v_1 \to v_2 \to v_3$ using $e_{12}$ and $e_{23}$, (b) or $v_1 \to v_3$ using only $e_{13}$.
% If we take the weight of the edges into account (i.e., the overhang length), we can see that one of these two paths carries less information than the other and therefore we can mark it as \emph{transitive} and remove it from the string graph.
Given two vertices $v_i$ and $v_j$, the edge $e_{i,j}$ between them defines an overlap between the two vertices (i.e., sequences). 
The weight $w_{i,j}$ of the edge $e_{i,j}$ stores the \emph{overhang length} or suffix length, i.e., the part of $v_j$ that goes beyond the overlap between $v_i$ and $v_j$, and its direction, i.e., how such sequences overlap with each other ( for example, $v_i$ and $v_j$ could have the same strand orientation and the final portion of $v_i$ could overlap with the initial portion of $v_j$).

If $G = (V, E)$ is a bidirected string graph, then a \textit{valid walk} in $G$ is a continuous sequence of edges where each vertex is entered by an inward head and exited by an outward head (unless it is the end of the path) or vice versa.

A contig is a collection of overlapping DNA sequences that together form a consensus region of DNA, such as a chromosome. 
To be considered a contig, a set of overlapping sequences must form a linear chain that is a valid walk.

\vspace{-.5em}
\section{Related Work}\label{sec:related}

Contig generation is an important step in any \emph{de novo} genome assembly pipeline, regardless of the type of sequencing technology (long-read or short-read). 
In this section, we summarize the literature on this kernel and explain how it relates to our work.

% Hifiasm
% As mentioned in Section~\ref{sec:back}, t
The paradigm for assembling long-read sequencing data is composed of three main stages: finding overlapping sequences to create an overlap graph, removing redundant information to simplify the graph, and generating the contig set.% that establishes consensus between sequences.

HiCanu~\cite{nurk2020hicanu} and Falcon-Unzip~\cite{chin2016phased} implement similar approaches inspired by the Bogart algorithm presented by Canu~\cite{koren2017canu} to generate the contig set in shared memory starting from a sparse long-read overlap graph or a string graph.
The Bogart module creates an assembly graph using a variant of the \emph{best overlap graph} strategy of Miller et al~\cite{miller2008aggressive}, where a \emph{best} overlap is the longest overlap to a given read end excluding \emph{contained} sequences (i.e., when all bases in one sequence are aligned to another sequence).
The Bogart algorithm removes overlapping sequences from the overlap graph to include only those that are within some tolerance of the global median error rate, and recalculates the longest overlapping sequences using only that subset (i.e., a sparse overlap graph). 
Bogart generates the initial contig set from the maximum non-branching paths in that graph.

The Shasta assembler~\cite{shafin2020nanopore} also uses a similar procedure by creating an undirected graph where each vertex is an oriented read (i.e., each read contributes two vertices to the read graph, one in its original orientation and one in the reverse complement orientation) and an undirected edge is created between two vertices when we find an alignment between the corresponding oriented sequences.
To reduce the high connectivity in the repeat regions, the Shasta assembler preserves only the $k$-nearest neighbor subset of the edges.
The contig set is created from the linear structures of the graph.

In contrast, Hifiasm~\cite{cheng2021haplotype} generates a primary assembly based on the topological structures of the graph and the phasing relationship between the different haplotypes using a bubble-popping procedure~\cite{li2016minimap}. 
Finally, a best overlap graph is used to deal with remaining unresolved substructures in the assembly graph.

A De Bruijn graph-based approach is common~\cite{medvedev2007computability} for genome assembly of short-read sequencing data, but it has not been suitable for long-read data in the past because of higher error rates.

% In genomics, a De Bruijn graph is a directed graph representing overlapping areas between nucleotide sequences.
% A vertex is a $k$-mer (i.e., a substring of fixed length $k$) originated from the input sequences, and the edge between two vertices stores the forward and backward extension of such a $k$-mer needed to form the adjacent $k$-mer. 
% For example, for $v_i =$ ACT and $v_j =$ CTT, the forward extension stored in the edge $v_i \to v_j$ is T, while the backward extension is A.

PaKman~\cite{ghosh2020pakman} is a parallel distributed memory contig generation algorithm for short-read sequencing technology.
PaKman introduces a new compact data representation of the De Bruijn graph, named PaK-Graph, which uses iterative compression to fit the graph into the memory available on each node. 
This compression enables low-cost replication of the graph across nodes, reducing the need for communication and creating an embarrassingly parallel procedure. 
PaKman also introduces an algorithm to perform non-redundant contig generation, i.e., to avoid two processes traversing the same path in the graph and generating the same contig.

 MetaHipMer~\cite{georganas2018extreme,hofmeyr2020terabase} and the earlier HipMer~\cite{georganas2015hipmer} are distributed-memory \emph{de novo} metagenome and genome assembly pipelines, respectively, designed for short-read data and thus also use the De Bruijn-based approach.
Both are implemented using a partitioned global address space model in either UPC~\cite{carlson1999introduction} or UPC++~\cite{bachan2019upc++}.
Contig generation is performed after the construction of a distributed hash table of $k$-mers with the left and right base extension from the input data stored with each $k$-mer. 
Each process then creates contigs by starting at a $k$-mer, walking left and right, appending the extension, and looking up the resulting $k$-mer in the hash table. 
These lookups often take place on remote nodes and are performed with either a remote memory operation or a remote procedure call. 
If a previously visited $k$-mer is reached during this process, the two contigs are merged.
Fine-grained synchronization prevents a data race that can occur when two processes attempt to merge at the same time.

% \Aydin{Give details of the HipMer approach of parallel contig generation}
% \Kathy{I tried to describe it succinctly above}

% \Aydin{How to reconcile this with the existence of Flye (below) and the claims that long reads are becoming less erroneous. Perhaps say that there are also attempts to make variations of de Bruijn graphs work for long reads, as exemplified by Flye, but the resulting algorithms are significantly more complex than algorithms used by OLC assembler and not suitable for scalable implementation on distributed-memory architectures due to their extremely fine-grained access patterns?}
As error rates decrease for long-read data, we find in the state of the art some attempt to use a De Bruijn graph approach for this type of sequencing technology, such as the shared-memory assembler Flye~\cite{kolmogorov2019assembly}.
Instead of generating a contig set, Flye first generates a \emph{disjointig} set, i.e. concatenating multiple disjoint genomic sequences, and then concatenates these error-prone disjointigs into a single string (in any order), constructs an assembly graph from the resulting concatenation, uses sequences to disentangle this graph, and resolves bridged repetitive areas (which are bridged by some sequences in the repeat graph).
It then uses the repeat graph to resolve unbridged repetitive areas ( that are not bridged by sequences) based on the differences between repeat copies. 
The output is a contig set generated from the paths in this graph.
Using a De Brujin approach in conjunction with long read sequencing data leads to more complex algorithms than those commonly used in an OLC-based assembler.
This makes an approach such as Flye's unsuitable for scalable implementation on distributed memory architectures, as the access patterns are extremely fine-grained.

\enlargethispage{-12pt}

\begin{figure}
    \centering
    \includegraphics[width=\columnwidth]{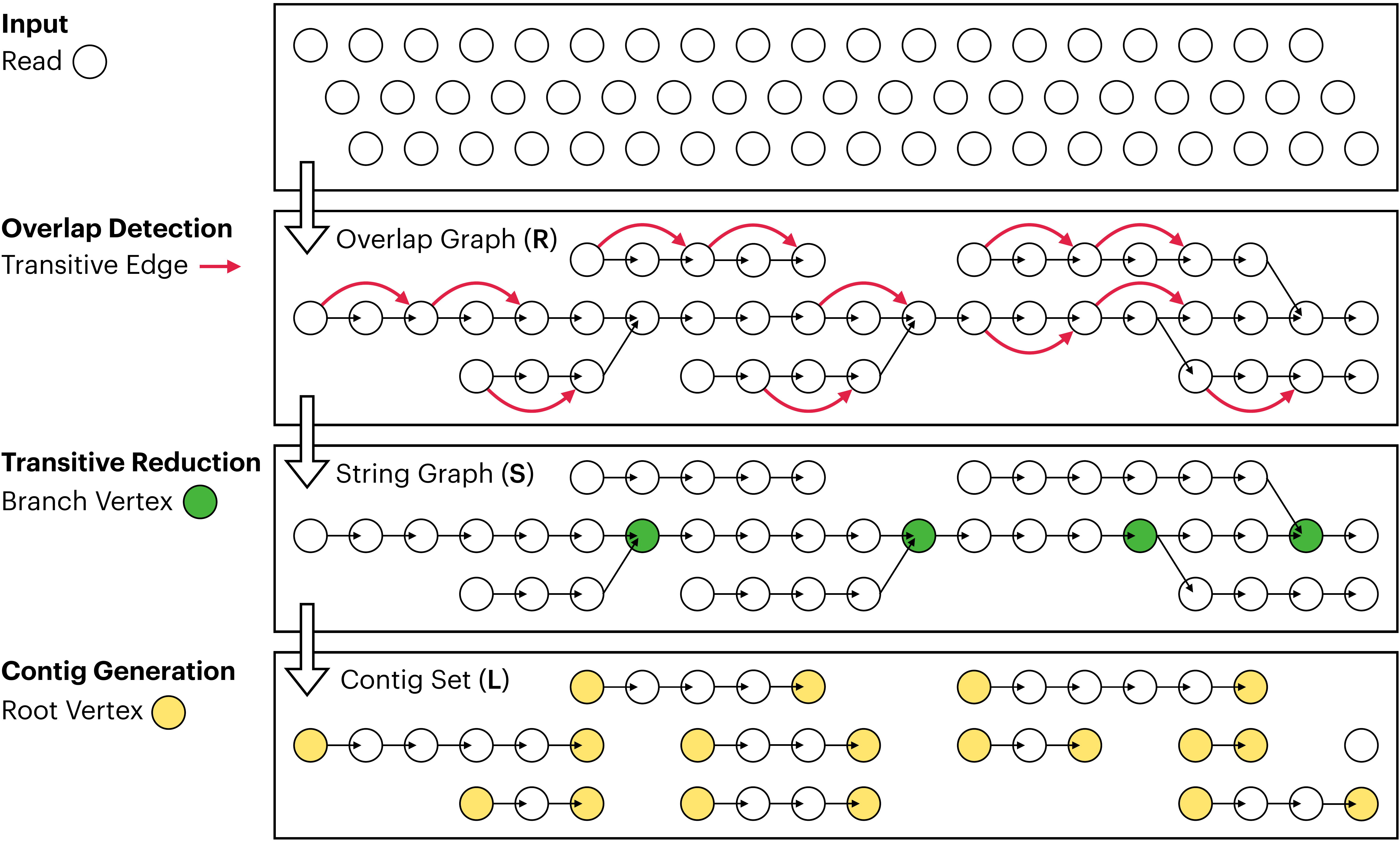}
    \caption{A high-level representation of ELBA built on diBELLA 2D~\cite{guidi2021parallel} and extended to include contig generation, which is critical to the assembly process.} \label{fig:pipeline}
%    \vspace{-1em}
\end{figure}

Regarding the use of sparse matrices and matrices in general for genomic computations, there is related work for both shared and distributed memory machines.
The shared-memory software BELLA~\cite{guidi2021bella} is the first work to propose the use of sparse matrices in the context of \emph{de novo} genome composition, focusing on the overlap detection phase.
A sparse matrix $\mA$ is used to indicate the presence of $k$-mers in sequences, and by multiplying by their transpose, i.e. $\mA\mA\transpose$, BELLA identifies overlapping sequences.

diBELLA 2D~\cite{guidi2021parallel} resembles BELLA, in that it computes both overlap detection and transitive reduction over an overlap graph as distributed SpGEMM and sparse computation.
PASTIS~\cite{selvitopi2020pastis}, similarly, computes protein homology search as distributed SpGEMM.

Besta et al.~\cite{besta2020communication} present an approach, similar to BELLA, for calculating Jaccard similarity between sequences of different genomes using distributed SpGEMM.
The main difference is that their software is optimized for the case where the output matrix is dense.

\vspace{-.5em}
\section{Algorithm}\label{sec:algortihm}

In this section, we introduce ELBA, a parallel distributed memory algorithm designed to generate a contig set from a string graph representation of long read sequencing data.

ELBA uses sparse matrices as the core data structure throughout the computation.
This abstraction leads to a flexible and modular software design, as well as high parallelization that performs well in distributed memory.
To implement parallel computation over sparse matrices, we use the Combinatorial BLAS (CombBLAS) library~\cite{CombBLAS20}, a well-known framework for implementing graph algorithms in the language of linear algebra, and use a semiring abstraction to overload the classical multiplication and addition operation as needed.
ELBA is inspired by Canu's best-overlap graph strategy~\cite{koren2017canu} and improves upon the transitive reduction algorithm presented in diBELLA 2D~\cite{guidi2021parallel} by extracting a set of high-quality, non-branching paths from the string graph.

Unlike MetaHipMer, which uses fine-grained communication at nearly every step of contig construction, ELBA localizes the graph traversal problem so that the sequences that make up a contig are stored locally at each rank, avoiding communication.

\subsection{Overview}

\begin{algorithm}[t]
    \caption{Parallel matrix-based computation in ELBA.}
    \label{alg:elba}
    \begin{algorithmic}[1] % The number tells where the line numbering should start
        \Procedure{ELBA}{}
             \State $\textit{sequences} \gets\Call{FastaReader}$ %\Comment{Read input}
             \State $\textit\kmer{s} \gets \Call{KmerCounter}$
             \State $\mA \hspace{0.43em}\gets \Call{GenerateA}{\textit{sequences},\hspace{0.0825em}\textit\kmer{s}}$
             \State $\mA\transpose \gets \Call{Transpose}{\mA}$
             \State $\mC \gets \mA\mA\transpose$\Comment{Candidate overlap matrix}
             \State $\mC \gets \Call{Apply}{\mC, \rm Alignment()}$\Comment{Run alignment}
             \State $\mR \gets \Call{Prune}{\mC, \rm AlignmentScoreLessThan(\textit{t})}$
             \State $\mR \gets \Call{Prune}{\mR, \rm IsContainedRead(\textit{})}$
             \State $\mS \gets \Call{TransitiveReduction}{\mR}$
            %  \State $\mv \gets \Call{ConnectedComponent}{\mR}$
            %  \State $\mpar \gets \Call{GreedyPartitioning}{\mv}$
            %  \State $\mL \gets \Call{InducedSubgraph}{\mR, \mpar}$
             \State $\textit{cset} \gets \Call{ContigGeneration}{\mS}$\Comment{Algorithm~\ref{alg:contig}}
            %  \State $\textit{genome} \gets \Call{Scaffold}{\textit{contigs}, \textit{reads}}$\Comment{Algorithm~\ref{alg:scaf}}
            \State \textbf{return} $\textit{cset}$
        \EndProcedure
    \end{algorithmic}
\end{algorithm}

%\input{pseudocodes/algorithm2}

% https://tex.stacfexchange.com/questions/115709/do-while-loop-in-pseudo-code

\begin{algorithm}[t]
    \caption{Parallel contig generation on $\mS$.}
    \label{alg:contig}
    \begin{algorithmic}[1] % The number tells where the line numbering should start
        \Procedure{ContigGeneration}{$\mS, sequences$}
            % \State $\Call{GetReadLength}{reads}$
            % \State $b \gets \Call{Init}{ } $\Comment{Branch vector}
            % \State $r \gets \Call{Init}{ } $\Comment{Root vector}
            % \State $n \gets \Call{Init}{ } $\Comment{Connected component size}
            % \State $dest \gets \Call{GetContigAssignment}{\mS, b, r, n}$\Comment{Algorithm~\ref{alg:assnmnt}}
            % \State $lidx \gets \Call{Init}{ } $\Comment{Local index vector}
            % \State $\mL \gets \Call{InduceSubgraph}{\mS, dest, lidx}$
            \State $\mL \gets \Call{BranchRemoval}{\mS}$
            \State $\mv \gets \Call{ConnectedComponent}{\mL}$
            \State $\mpar \gets \Call{GreedyPartitioning}{\mv, P}$ %\Comment{Algorithm~\ref{alg:partitioning}}
            \State $\mP \gets \Call{InducedSubgraph}{\mL, \mpar}$%\Comment{Algorithm~\ref{alg:induced}}
            \State $\textit{cset} \gets \Call{LocalAssembly}{\mP, sequences}$
            \State \textbf{return} $\textit{cset}$
        \EndProcedure
    \end{algorithmic}
    
\end{algorithm}

ELBA uses a highly parallelizable strategy to generate the contig set in a distributed memory environment. This means that each processor may not have access to the entire set of sequences it needs for contig generation because they are distributed among processes.
Read sequences must therefore be communicated across the processor grid before contigs are generated via depth-first search.

To efficiently communicate sequences where they are needed, ELBA uses a parallel algorithm, implemented as a sparse matrix-based computation over the string matrix, to extract information about which sequences belong to the same contig (i.e., a \emph{linear component}) and therefore need to be sent to the same processor.
A linear component of a graph is the maximal subgraph where each vertex is connected to two adjacent vertices, except for two vertices that mark the extremities of the linear chain, which are connected to a vertex.
Once the membership of each sequence is known (i.e., to which contig it belongs), ELBA uses this information to achieve load balancing that ensures that each processor has similar amount of work during the local assembly process.

This is achieved by first estimating the contig sizes, i.e., the number of reads belonging to a particular contig, in parallel and then assigning approximately equal collections of contig sets to each process.
Then, the sequences are redistributed among the processes using a newly implemented function that is also based on sparse matrix computation. This function generates local matrices from a distributed matrix (in our case, the string matrix) on each process according to the load-balanced assignment. 
ELBA then assembles the contigs whose sequence connectivity is stored in the local matrix in parallel on each processor without requiring any further communication.

Algorithm~\ref{alg:elba} summarizes the entire computation, including overlap detection and transitive reduction---lines 11-12 represent the main contribution of this work, better represented in Algorithm~\ref{alg:contig}: (a) how the contig set is determined (line 2), (b) how the workload is distributed among the processes (lines 3-4), and (c) how the local assembly is performed (line 5) and the contig set is output.

\subsection{Contig Set Determination}

The first step in our contig generation algorithm is to determine which sequences belong to which contig based on the connectivity of the string graph.

Let us define a linear component of the graph as the maximal subgraph $L \subseteq R=(V,E)$ where (a) each vertex has either degree one or two and (b) the subgraph is connected. 
In a linear component there are exactly two vertices which can have degree one due to the connectedness rule. 
These vertices are the endpoint vertices.

Then, we define the \emph{contig set} of $R$ as the set of all linear chains which are not themselves subgraphs of another linear chain in $R$.
Thus, the contig set consists of several independent collections of nodes in the bidirected string graph, where each collection is a linear sequence of unique overlapping sequences. 
The determination of the contig set can be done in two stages.

First we have to identify \emph{branching vertices} that are vertices whose degree is $\geq$ 3. 
These vertices would make it impossible to determine a unique linear chain. 
Therefore, our goal is to mask out such branching vertices to obtain only linear sequences of vertices. 
For example, consider a string graph consisting of (a) $v_1 \to v_2 \to v_3$, (b) $v_3 \to v_4 \to v_5 \to v_6$, and (c) $v_3 \to v_7 \to v_8$, the vertex $v_3$ is a branching vertex since its degree is equal to three, and it would not allow us to extract a linear chain, since at this point in the computation we cannot know whether the correct linear path is (i) $v_1 \to v_2 \to v_3 \to v_4 \to v_5 \to v_6$ or (ii) $v_1 \to v_2 \to v_3 \to v_7 \to v_8$.
Thus, we mask out $v_3$ to obtain three linear sequences of vertices: (i) $v_1 \to v_2$, (ii) $v_4 \to v_5 \to v_6$, and (iii) $v_7 \to v_8$.

To identify branching vertices, we perform a summation reduction over the row dimension of the string matrix (i.e., an adjacency matrix) and return a distributed vector $\md$ whose values represent the degree of the corresponding row sequence (i.e., the index of the vector).
Then we perform an element-wise selection operation on the degree vector $\md$ to extract the indices (sequences) whose value is greater than or equal to 3. 
The result is a distributed vector $\mb$ whose values are the indices of the branching vertices.
The branch vector $\mb$ is used to remove the corresponding rows and columns from the string matrix $\mS$ and create a linear chain version of $\mS$ that we name $\mL$ (line 2, Algorithm~\ref{alg:contig}).
From a graph-theoretic point of view, this operation removes (i.e., sets to zero) the edges adjacent to branching vertices. 
The indexing of the matrix does not change, but its nonzeros do. 
For example, if row 10 is a branching vertex, the entire row---and column, since $\mS$ is symmetric---is cleared, but row 10 is still a row in the matrix. 
This way we can avoid re-indexing sequences during the computation.
The result is an even sparser string graph with nodes of degree $0$, $1$ or $2$. 
For each contig, there are exactly two vertices of degree $1$, which we call root nodes (or root vertices) and use as starting points for depth-first search and local assembly of the contig in the final stage of contig generation.

Once the string matrix is in its unbranched form $\mL$, we want to decompose it into its linear components to produce manageable independent subproblems that we can work on in parallel (i.e., local assemblies).
Therefore, we use the sparse matrix based connected components (LACC) algorithm presented by Azad et al.~\cite{lacc2019} to determine the contig set.
LACC is a distributed-memory implementation of the Awerbuch-Shiloach algorithm~\cite{awerbuch1987new} using the CombBLAS library.
LACC takes advantage of the sparsity of the vector to avoid processing inactive vertices.
It takes as input the unbranched sparse matrix $\mL$ and returns a distributed vector $\mv$ (line 3, Algorithm~\ref{alg:contig}), which is a mapping from global sequence indices (i.e., rows and columns of $\mS$) to contig indices $C_i$.
In the previous graph example, LACC would return ${v_1, v_2} \in C_1$, ${v_4, v_5, v_6} \in C_2$, and ${v_7, v_8} \in C_3$.

Once we have determined which sequences belong to which contig, we estimate the size of each contig by counting how many sequences belong to it.
However, since $\mv$ is a distributed vector, a processor may not know the full set of sequences belonging to the contig it owns. 
Therefore, we need to communicate this information across the processor grid.
Each processor computes a local size estimate based on the vertices it owns locally for each local contig.
An MPI Reduce-scatter collective operation is used to determine the global size of each contig and redistribute the sizes across the processor grid to create a distributed mapping of contig indices to their associated global sizes.

\subsection{Contig Load Balancing and Communication}

Previously, we determined which sequences belong to which contig and estimated the size of each contig based on the number of sequences associated with it. %(this information is stored in the distributed vector $\mv$)
Using the contig size as an \nobreak optimization parameter, we want to distribute the workload as evenly as possible across the processor grid.

\paragraph{Load Balancing Algorithm} 
Given a vector of contig sizes of length $n$ and $P$ processes, we want to find the near-optimal contig-to-processor assignment such that the amount of work estimated by the contig size that each processor has to do is similar.
Our objective and problem definition is similar to the \emph{multiway number partitioning} optimization problem first introduced by Ronald Graham in the context of the identical-machines scheduling problem~\cite{graham1966bounds}.
The classical application is to schedule a set of $m$ jobs with different runtimes on $k$ identical machines in such a way that the makespan, i.e., the elapsed time until the schedule is completed, is minimized.

In the context of contig generation, this means that given a multiset of $S$ instances (in our case, the contig sizes), we want to partition this multiset into $P$ subsets (the number of processes in our processor grid and must be a positive integer) such that the sums of the subsets (i.e., the sums of the contig sizes) are as similar as possible.
The partitioning results in a balanced distribution of the workload for the next and final phase of the computation.

The multiway number partitioning problem is NP-hard.
To overcome this limitation, in ELBA we use an approximation algorithm known in the scheduling literature as the Longest Processing Time (LPT) algorithm, whose goal is to minimize the largest subset, which belongs to a class of algorithms known as \emph{greedy number partitioning}.
The algorithm loops over the contig sizes and inserts each number into the set whose current sum of sizes is smallest.
The result is a partitioning that minimizes the time processes spend waiting for the most heavily loaded process to finish assembling its contig subset.
If the contig sizes were not sorted, then the runtime would be $O(n)$ and the approximation ratio would be at most $2-1/P$. 
It is possible to improve the approximation ratio to $(4P-1)/3P$ by sorting the input vector of contig sizes~\cite{graham1969bounds}.

The improvement in the approximation using LPT is accompanied by an increase in the runtime to $O(n \textrm{log} n)$ due to sorting.
However, the number of contigs $n$ is smaller than the number of sequences by at least an order of magnitude, and the increase in runtime does not create a computational bottleneck.
For the same reason, we collect the global information about contig lengths in a single processor and run the partitioning algorithm on it to avoid the unnecessary communication of small messages.
The partitioning algorithm returns the vector $\mpar$ specifying the assignment of contigs to processes (line 4, Algorithm~\ref{alg:contig})---$\mpar$ is broadcasted to the entire processor grid so that each local process can determine where to send its local sequences and associated information.

As mentioned earlier, the problem size at this stage of the computation is often smaller than the problem size at the initial stage (i.e., overlap detection), so for some species it is possible that $n < P$.
In this case, some of the processes are idle for the final phase of the computation.
For the two species we use in the experimental evaluation in Section~\ref{sec:soa}, $n$ is equal to $6411$ and $4287$ and $P$ varies from $18$ to $128$.
In the next section, we explain how contigs are redistributed among processes based on their size.

\paragraph{Induced Subgraph Algorithm} 

Once we have determined where a contig and its sequences should be stored, we must perform the \nobreak actual communication step to send the linear component information and associated sequences to the owner processor.

Communicating both the linear component information and the sequences involves the same high-level procedure of reassigning vertices representing a linear chain to their owner processor.
However, the underlying data structures are fundamentally different, namely that the overlap graph is stored as a sparse matrix while the sequences are stored as distributed char arrays, and therefore require a different implementation.

Let us first focus on the communication of linear component information, i.e. the graph-like structure that stores connectivity information. We have as input the sequence-by-sequence matrix $\mL$, the resulting matrix after the transitive reduction step and from which we have cut out vertices of degree $\geq 3$.
$\mL$ is distributed over $P$ processes, which are logically organized in a $\sqrt{P} \times \sqrt{P}$ grid. %, i.e., each process has a subset of rows and columns. 
Let $n$ be the number of vertices in $\mL$, where $\mL = (V, E)$ in the graph interpretation. 
From the load balancing algorithm, we have a distributed vector $\mv : [n] \to [P]$ such that $\mv[u] = P_i$ means that vertex $u$ should belong to processor $P_i$.

The goal is to create an induced subgraph---or submatrix---$\mL^{(P_i)}$ locally on each processor $P_i$, i.e., a graph formed by a subset of the vertices of the original graph $\mL$ and the edges connecting the vertices in this subset.
Formally, we define an induced subgraph $\mL^{(P_i)}$ such that $\mL^{(P_i)} = (V^{(P_i)}, E^{(P_i)})$ with $V^{(P_i)} = \{v \in V : P_i = \mv[v]\}$ and $E^{(P_i)} = \{(u,v) \in E : u,v \in V^{(P_i)} \}.$

The vector $\mv$ is also distributed across the $\sqrt{P} \times \sqrt{P}$ processor grid and is therefore divided into $P$ subvectors, each of size $\approx n/P$; note that we write $\mv_{(i,j)}$ to denote the subvector on process $P(i,j)$. 
Each process is only aware of which vertices it stores to send to other processes. Therefore, we need a communication step to make each process aware of which vertices to receive.
Given the way $\mv$ is distributed, we can avoid an MPI\_Allgather operation spanning the entire grid and instead use the square process grid to communicate $\mv$ in a scalable way.
That is, we perform an allgather operation over the $\mathbf{Row}$ dimension followed by point-to-point communication to access the information stored on the $\mathbf{Column}$ dimension.

\begin{figure}
    \centering
    \includegraphics[width=\columnwidth]{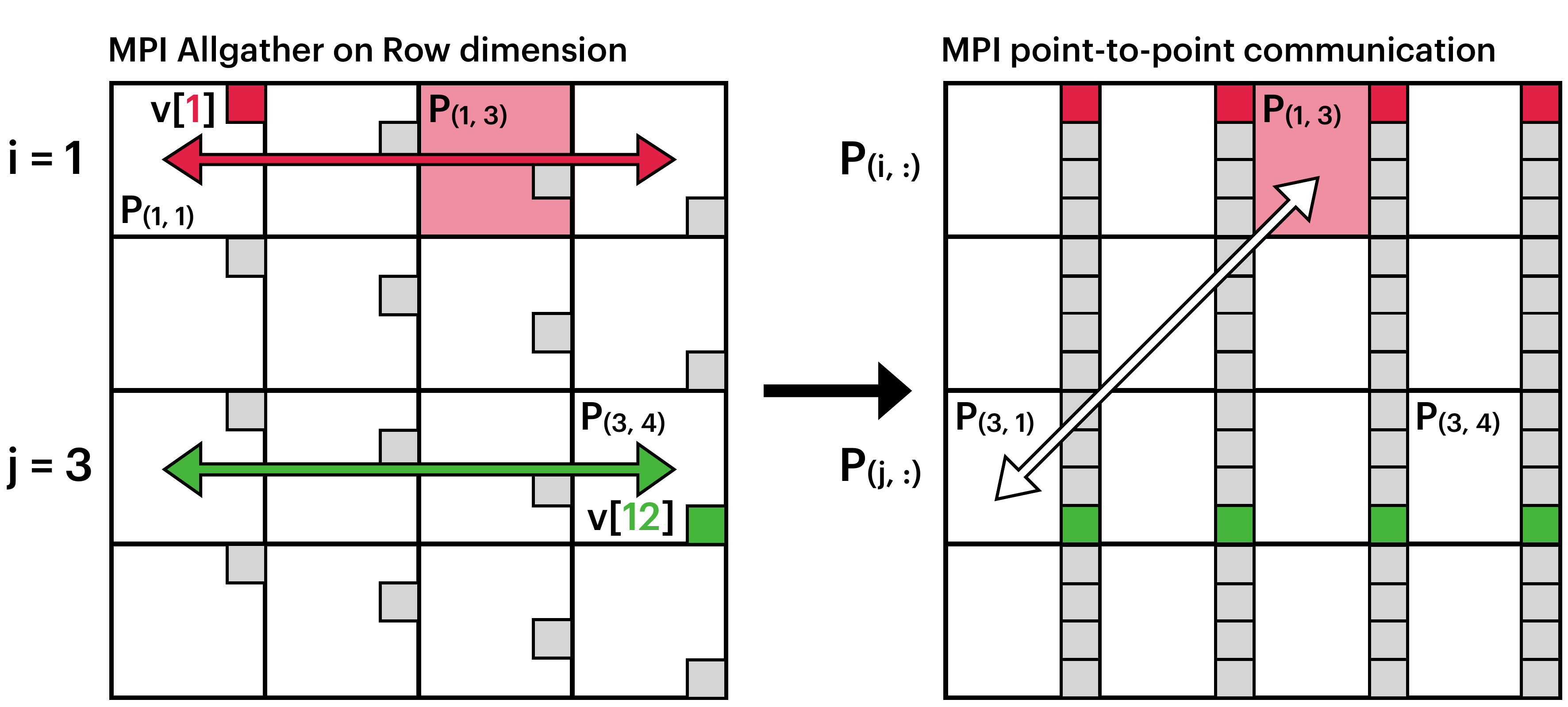}
    \caption{An example of how the induced subgraph algorithm communicates vertices across the processor grid.}
    \label{fig:induced}
    \vspace{-1em}
\end{figure}

More precisely, let $(u,v) \in \mL_{i,j}$ be any nonzero value, where $\mL_{i,j}$ is the local submatrix stored on processor $P(i,j)$. 
The goal is to determine $\mv[u]$ and $\mv[v]$. 
Given we know $(u,v)$ is stored in the submatrix $\mL_{i,j}$, we know that $\mv[u]$ is stored in the row $P(i,:)$. 
Thus, we can access $\mv[u]$ by computing an MPI\_Allgather operation over the processor row $P(i,:)$ and, more generally, over the $\mathbf{Row}$ dimension of $\mS$.
To determine $\mv[v]$, we first note that if we are a diagonal processor, i.e. $i = j$, we already have access to $\mv[v]$ via the previous MPI\_Allgather operation. 
If $i \neq j$, then we know that $\mv[v]$ is stored on the transposed processor $P(j,i)$ because we performed the allgather operation over the $\mathbf{Row}$ dimension, i.e. $P(j,:)$.
Therefore, we perform a point-to-point communication between $P(i,j)$ and $P(j,i)$ to exchange the subvector $\mv_{i,*}$ and the subvector $\mv_{j,*}$. 
This gives the processor $P(i,j)$ access to $(u,v)$. 
It follows that each processor $P(i,j)$ now has access to every entry in $\mv$ corresponding to a nonzero in $\mL_{i,j}$. A similar procedure is previously implemented for distributed-memory breadth-first search~\cite{bulucc2011parallel}.

Figure~\ref{fig:induced} illustrates an example with 1-based indexing, where $u = 1$ (red square in the leftmost matrix) is stored on process $P(1,1)$ and $v = 12$ (green square) is stored on process $P(3,4)$.
If we assume that $P(1,3)$ requires access to $\mv [1]$ and $\mv [12]$,
an MPI\_Allgather operation on the $P(1,:)$ row would provide access to $\mv [1]$, while the same operation on the $P(3,:)$ row would provide access to $\mv [12]$ on the transposed processor $P(3,1)$.
Point-to-point communication between $P(1,3)$ and $P(3,1)$ would then allow $P(1,3)$ to access $\mv [12]$.

Once we have access to the processor assignment stored in $\mv$, we communicate edges to their target processor and build the local induced subgraph. % from such edges.
To do this, we loop through each nonzero $(u,v) \in \mL_{i,j}$ stored on processor $P(i,j)$, and for each $(u,v)$ where $\mv[u]$ and $\mv[v]$ are destined for the same destination processor, we construct a triple $(u,v,\mS(u,v))$ and place it on an outgoing buffer to the destination processor. 
A custom all-to-all ensures the required re-distribution of non-zeros.
Once each processor has access to its edge set $E^{(P_i)}$, the local adjacency matrix 
of the induced subgraph is constructed; while we re-index the local matrix to fit its new, smaller size, we also keep a map of the original global vertex indices, since it is needed in the final phase.

\paragraph{Read Sequence Communication} 

The communication of read sequences is implemented separately, since the read sequences are not stored as nonzeros in the sparse matrix $\mL$, but in a distributed auxiliary data structure. 
Furthermore, a sequence is represented and stored as a char array. A large data set could exceed the MPI limit of $2^{31}-1$. 
Therefore, we need to treat sequence communication separately and consider this potential limitation.

Read sequences that need to be sent to a processor other than the one they are currently on are packed into a char buffer and communicated as a sequence of non-blocking point-to-point messages in an all-to-all fashion.
To deal with the MPI $2^{31}-1$ count limit, we check the length of each message to be communicated and its receive buffer. 
If it goes beyond the limit, we communicate the sequences using a user-defined contiguous MPI data type whose size is equal to the buffer length. 
This way we can send and receive each character buffer in a single call.

\subsection{Local Contig Assembly}

Let $\mL_{i,j}$ be the local graph (or matrix), composed of one or more linear components, stored on the processor $P_{i,j}$ obtained via the induced subgraph algorithm. 
There are $P$ such graphs, one for each processor, but we can assume without loss of generality that we are dealing only with $\mL(P_{i,j})$ or $\mL$ for short.

Suppose $\mL$ has $n$ vertices and $m$ edges. 
Each vertex of $\mL$ is a read sequence $l$ assigned by the partitioning algorithm to $P_{i, j}$. 
Then, we define $\Sigma = \{A, C, T, G\}$ as the alphabet of DNA nucleotides. 
Consequently, we represent the $n$ sequences of $\mL$ by $l_0, l_1, \ldots, l_{n-1} \in \Sigma^{*}$, which means that each sequence is a combination of $\{A, C, T, G\}$.
Given any read sequence $l$, we express the nucleotides of that sequence by $(l[0], l[1], \ldots, l[|l|-1])$, where $|l|$ denotes the length of the sequence.
If $l[i]$ is a base or nucleotide in $\Sigma$, we denote the Watson-Crick complement base of $l[i]$ by $l[i]^c$. 
This allows us to generalize the notion of sequence to include its reverse complement, as defined in Section~\ref{sec:back}: if $i < j$, then we write $l[i:j]$ for the substring $(l[i], l[i+1], \ldots l[j])$ and $l[j:i]$ for its reverse complement substring $(l[j]^c, [j-1]^c, \ldots, l[i]^c)$. 

$\mL$ can also be viewed from the point of view of its matrix representation, such that $\mL$ is a sparse $(n \times n)$ matrix with $m$ nonzeros. 
In the earlier stages of the pipeline, we use the doubly compressed sparse column (DCSC) format~\cite{buluc2008representation} to store our matrices for scalability. 
However, for the local traversal algorithm described in this section, we converted these matrices to compressed sparse column (CSC) format for simplicity and faster vertex (column) indexing. The storage space required to store the local matrices is an order of magnitude smaller than before and hence CSC does not introduce memory scalability issues despite hypersparsity.
% Since the local assembly takes relatively little time compared to the earlier stages of the computation, 
This conversion takes linear time in the number of local vertices, as only column pointers needs to be uncompressed and row indices array stays intact, and has negligible effect to overall runtime.

As a sparse matrix, $\mL(u,v) = e$ is a nonzero provided that the source sequence $l_u$ and the destination sequence $l_v$ overlap. 
In this nonzero $e$ we store two values computed from the original string graph $\mS$, which is all we need for assembly: $\mpre_u(e)$ and $\mpost_v(e)$. 
To define $\mpre_u(e)$ and $\mpost_v(e)$, we use inclusive indexing, resulting in the following asymmetric definition.
The first value $\mpre_u(e)$ stores the index $i$ of $l_u$, which is the last nucleotide in $l_u$ that does not overlap with $l_v$, i.e., the nucleotide on $l_u$ that precedes the overlap with $l_v$. 
The second value $\mpost_v(e)$ stores the index $j$ of $l_v$, which is the first nucleotide in $l_v$ that  overlaps with $l_u$, i.e., the index of the beginning of the overlap between $l_u$ and $l_v$.
The definition is asymmetric because $l_u[i]$ and $l_v[j]$ do not store the same nucleotide.
This asymmetry is necessary to compute the non-overlapping prefixes of each read in a contig, which, when joined together, produce a contig sequence.
% \Gabe{
%The first value $\mpre_u(e)$ stores the position of $l_u$ or index $i$ of $l_u[i]$ where the overlap region between $l_u$ and $l_v$ begins on $l_u$, while the second value $\mpost_v(e)$ stores the index of the beginning of the overlap region between $l_u$ and $l_v$ on $l_v$.
% This asymmetry is due to the usage of inclusive indexing.
%We note that $\mpre$ values denote indices one nucleotide before an overlap, while $\mpost$ values denote indices beginning an overlap. 
% }
% \Aydin{I feel like these concepts are not symmetric. $\mpre_u(e)$ is the index "one before" the actual overlap on $l_u$ whereas $\mpost_v(e)$ is the actual overlap on $l_v$. In other words, $\mpre_u(e)$ is the index of the last nucleotide that does not overlap while $\mpost_v(e)$ is the index of the first nucleotide that overlaps. The current definition reads as if they are symmetric but looking at Fig 3 and the following paragraph, it is clear they are not symmetric. I think the definition should reflect that asymmetry and make it clear.}. 

\begin{figure}
    \centering
    \includegraphics[width=\columnwidth]{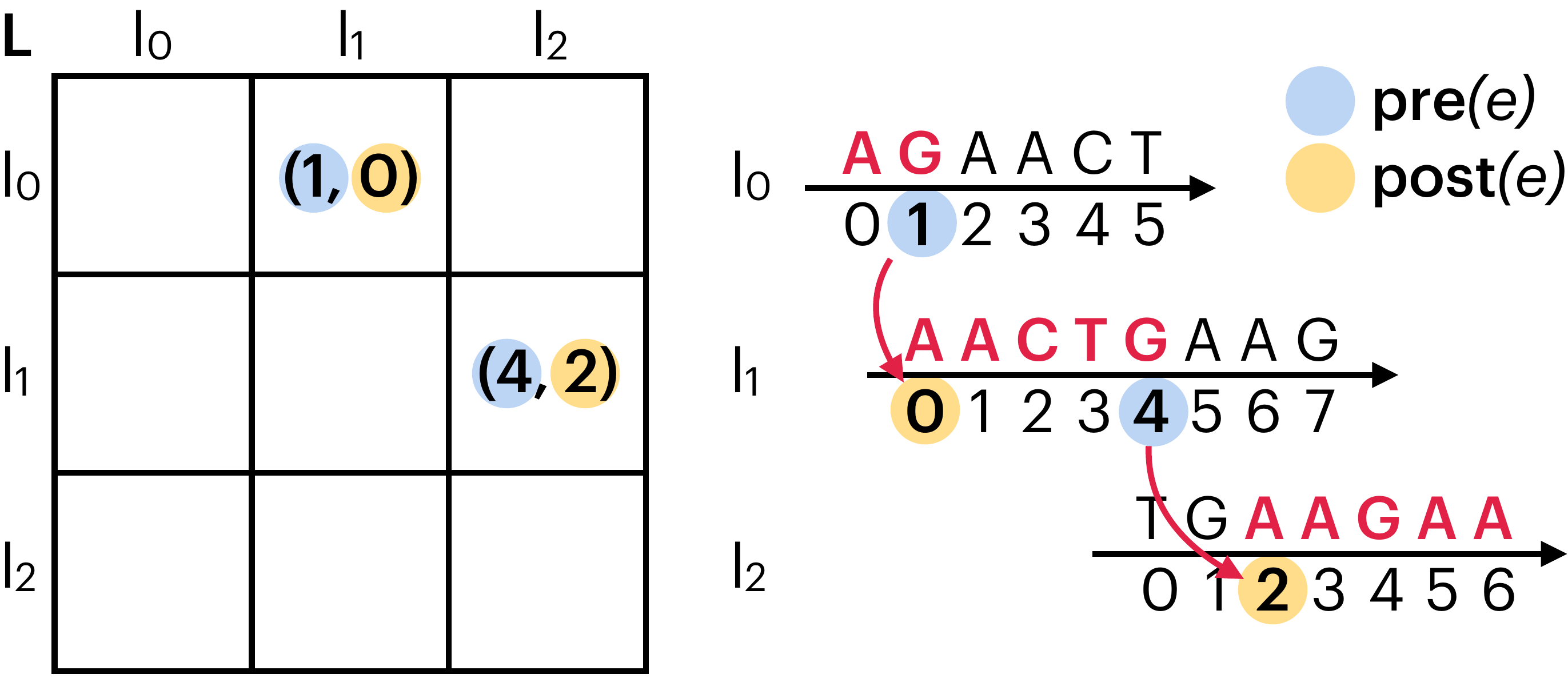}
    \caption{An example of how the local assembly algorithm concatenates sequences.}
    \label{fig:locassembly}
    \vspace{-1em}
\end{figure}

For example, if $l_u = l_0 = AGAACT$ and $l_v = l_1 = AACTGAAG$ as shown in Figure~\ref{fig:locassembly} and we consider 0-based indexing, then $\mpre(e) = 1$ and $\mpost(e) = 0$ because $l_u[2 : 5]$ and $l_v[0 : 3]$ overlap. 
If instead we consider $l_u = l_1 = AACTGAAG$ and $l_v = l_2 = TGAAGAA$, then $\mpre(e) = 4$ and $\mpost(e) = 2$ because $l_u[5 : 7]$ and $l_v[2 : 4]$ overlap.
It is also possible that $l_u = l_0$ would overlap with $l_v^c = l_1^c = CTTCAGTT$ (the reverse complement of $l_1$). In this case, the same procedure would apply to compute $\mpre(e) = 1$ and $\mpost(e) = 4$, where the overlapping subsequences would be $l_u[2 : 5]$ and $l_v^c[7 : 4]$, since the $l_u$ subsequence $AACT$ would overlap with the $l_v^c$ reserve complement subsequence $AGTT$.

%The reader may wonder why it is necessary to store the value $\mpost(e)$, 
%given that in a theoretical scenario the beginning of the overlapping substring of $l_v$ should always be $0$ or $|l_v| - 1$, depending on the orientation.
Theoretically, the beginning of the overlapping substring of $l_v$ should always be $0$ or $|l_v| - 1$, depending on the orientation. Yet, we still store $\mpost(e)$ because the diBELLA 2D~\cite{guidi2021parallel} pipeline, which we extend in this work, uses an x-drop seed-and-extend approach to align overlapping sequences. 
With x-drop, the alignment between two sequences can potentially end early (i.e., before extending to the end of a read), leaving a short overhang in the alignment coordinates at the end of the sequence. 
If $l_u = l_1 = AACTGAAG$ and $l_v = l_2 = TGAAGAA$ are defined as in Figure~\ref{fig:locassembly}, and the alignment told us that $l_u[5 : 7]$ and $l_v[2 : 4]$ were the overlapping region, it would be incorrect to match $\mpre(e) = 4$ with $\mpost(e) = 0$, since we need $\mpost(e) = 2$ to correctly concatenate the subsequences that form a contig.
If we join the three partial sequences marked in red on $l_0, l_1, l_2$ in Figure~\ref{fig:locassembly}, as explained below, we get a contig.

%\input{tables/table2-machines}

% \usepackage{booktabs}
\begin{table*}[t]
\caption{Details of the machines used for evaluation; the frequency reported is the processor max turbo frequency.}
%: name, number of physical cores per node, processor max turbo frequency, processor model, memory, network, and L1, L2, L3 caches sizes.
% }
% \begin{adjustbox}{width=\textwidth}
% {
\centering
\resizebox{\textwidth}{!}{
\begin{tabular}{|l|c|c|l|c|c|r|r|r|}
\hline
\textbf{Platform} & \textbf{Cores/Node} & \textbf{Frequency (GHz)} &
%\textbf{N$\nicefrac{1}{2}$} &
\textbf{Processor} & \textbf{Memory (GB)} & \textbf{Network and Topology} & \textbf{L1}\hspace{.65em} & \textbf{L2}\hspace{.65em}  & \textbf{L3}\hspace{.65em}  \\ 
\hline
\hline
Cori Haswell & 32 & 
3.6 & Intel Xeon E5-2698V3 & 128 & Aries Dragonfly & 64KB & 256KB & 40MB \\
\rowcolor{Gray}
% Cori KNL & 68 & 1.6 & Intel Xeon Phi 7250  & 96 & Aries Dragonfly & 64KB & 1MB & - \\
Summit CPU & 42 &  4.0 & IBM POWER9  & 512 & InfiniBand Non-Blocking Fat Tree  & 32KB & 512KB & 10MB\\[-11pt]
% AWS c5n.18xlarge & 36 & 3.5 & Intel Xeon Platinum 8000 &  206 & Elastic Fabric Adapter (EFA) & & &  \\
&&&&&&&&\\
\hline
\end{tabular}}
\label{tab:machines}
\end{table*}

%\input{tables/table1-dataset}

% \usepackage{booktabs}
\begin{table}[t]
\caption{
% Details about the d
Data sets used during evaluation: name, depth, number of sequences in the input, average read length, input size, genome size, and error rate.
}
\centering
\resizebox{\columnwidth}{!}{
\begin{tabular}{|l|c|r|r|c|r|c|}
\hline
\textbf{Label} &  \textbf{Depth} & \textbf{Reads (K)} & \textbf{Length} & \textbf{Input (GB)} & \textbf{Size (Mb)}  & \textbf{Error} (\%) \\ 
\hline
\hline
% H. sapiens & \hspace{0.5em}10 & 4,421.6 & 7,402 & 31.6 & 3,000 & 15.0 \\
O. sativa & \hspace{0.5em}30 & 638.2 & 19,695 & 12.2 & 500 &  \hspace{0.5em}0.5 \\
\rowcolor{Gray}
C. elegans & \hspace{0.5em}40 & 420.7 & 14,550 & \hspace{0.5em}3.8 & 100 &  \hspace{0.5em}0.5 \\
H. sapiens & \hspace{0.5em}10 & 4,421.6 & 7,401 & 31.1 & 3,200 &  15.0 \\
% \rowcolor{Gray}
% S. cerevisiae & 383 & 227.3 & 20,166 &  \hspace{0.5em}4.4 & 12 &  \hspace{0.5em}0.5 \\
% E. coli & 302 & 95.6 & 14,548 & \hspace{0.5em}1.3 & 5 &   \hspace{0.5em}0.2 \\
\hline
\end{tabular}}
\vspace{-1em}
\label{tab:data}
\end{table}

Because the local assembly operates directly on the CSC format, we describe it briefly. 
$\mL$.JC is the column pointer array of length $n+1$, $\mL.IR$ is the row index array of length $m$, and $\mL.\textit{VAL}$ is the array of tuples $(\mpre(e), \mpost(e))$, also of length $m$.

The local assembly algorithm is a variant of depth-first search, simplified by the fact that the maximum vertex degree of $\mL$ is $2$ by construction. 
For this reason, there is always only one vertex in the frontier, and the search is thus a linear walk. 
Each local matrix $\mL_{i,j}$ is assigned a contig set by multiway number partitioning, which is the set of connected components of $\mL_{i,j}$. 
Each vertex of $\mL$ has degree $1$ or $2$. 
Moreover, we define contigs as linear chains of at least two sequences, where $q$ is the number of vertices in a given connected component.
It follows that any connected component consisting of $q \geq 2$ vertices must have exactly $2$ vertices of degree $1$ (i.e., \emph{root} vertices) and $q-2$ vertices of degree $2$ (i.e., \emph{intermediate} vertices).
% \Giulia{Let's define $q$ properly.}
% \Gabe{$q$ is just a number, the number of vertices in a given connected component, How can it be defined more specifically? like saying: I have $n$ apples and $m$ pears etc.}
% \Giulia{what you just wrote is not what you said in the main text.}
In short, the idea is to scan for root vertices in $\mL$ and, if we find them, to take a walk from that root vertex (via as many intermediate vertices as necessary) until another root vertex is found. 
As we proceed, we perform the concatenation of subsequences described below. 
This search for the root vertex is performed over all $n$ vertices. 
Therefore, we must mark each final root vertex (found by the linear traversal) as visited to avoid accidentally composing the same contig twice.
% \Giulia{Checkpoint.}

% not needed
% \input{pseudocodes/algorithm5}

% \Giulia{we loop over what?}
% \Gabe{like a for loop: for (int i = 0; i <= n - 1; ++i)}lk
More precisely, we loop over the $n$ sequence vertices in each subgraph $\mL_{i,j}$ and compute $\mL_{i,j}.JC[i+1] - \mL_{i,j}.JC[i]$ from the column pointer array representing the degree of vertex $i$. 
Each time we find a vertex of degree $1$, i.e., a root vertex, that has not been visited before, we perform a traversal starting from this that root vertex $r$. 
Given a generic vertex in the chain $c$ (which can be either a root vertex or an intermediate vertex), we find the successor vertex $c'$ by examining the vertices in the slice $\mL_{i,j}.IR[\mL_{i,j}.JC[c] : \mL_{i,j}.JC[c + 1]]$ of the row index array where the edges are stored. 
Each vertex $c$ has at most two successor vertices, and we select the unvisited one.
This process continues until we reach the vertex $r'$ for which $\mL_{i,j}.JC[r' + 1] - \mL_{i,j}.JC[r'] = 1$, i.e., the second root vertex of a contig. 
The result is a chain of $q$ vertices $r, c_1, \ldots, c_{q-2}, r'$ for each contig stored in the submatrix $\mL_{i,j}$. 
As we proceed, we collect the edges between them $e_0, e_1, \ldots, e_{q-2}$.
% We continue this process until we reach the vertex $r'$, for which $\mL.JC[r' + 1] - \mL.JC[r'] = 1$. 
% The result is a chain of $q$ vertices $r, c_1, \ldots, c_{q-2}, r'$. Along the way, we collect the edges between them $e_0, e_1, \ldots, e_{q-2}$.
Because the values for $\mpre(e_i)$ and $\mpost(e_i)$ have already been computed and stored in each edge $e_i$, we know exactly which subsequences of which read sequence to look up and join to form the contig. 
Namely, $l_{r}[\alpha : \mpre(e_0)] \oplus l_{c_1}[\mpost(e_0) : \mpre(e_1)] \oplus \ldots \oplus l_{c_{q-2}}[\mpost(e_{q-3}) : \mpre(e_{q-2})] \oplus l_{c_{r'}}[\mpost(e_{q-2}) : \beta],$ where $\alpha = 0$ or $|l_{r}|-1$ depending on the orientation of $l_r$, and $\beta$ is similarly defined for $l_{r'}$.

This algorithm is performed by each process in parallel on its own induced subgraph $\mL_{i,j}$. 
For each read in the contig, we either look for the subsequence in the locally stored char array that we started with, or in the char array obtained by communicating the read sequence. 
Because we store read sequences in large packed char arrays, we do not need to copy the entire read sequence to find the correct subsequence. 
Instead, we can simply use the offsets already computed, which tell us where each read is in the buffer, and then read the subsequence directly from the buffer. 
The algorithm is $O(q)$, where $q$ is the number of vertices in a connected component as previously defined (i.e., the number of sequences in a contig), since the search for the root vertex takes linear time and the traversal is linear in the number of edges, which is $2(q-1)$.

\vspace{-.5em}
\section{Experimental Setup}

%To evaluate the performance of the algorithm presented in this paper and the performance of the \emph{de novo} long-read assembly pipeline, ELBA, into which we integrated our work, 
To evaluate our contig generation algorithms and the ELBA long-read assembly pipeline we used two machines: the Haswell partition of the Cray XC40 supercomputer Cori at NERSC and the IBM supercomputer Summit at Oak Ridge National Laboratory, on which we used only IBM POWER9 CPUs (Table~\ref{tab:machines}).

Each node on the Cori Haswell partition is a dual-socket 2.3GHz Intel Xeon with 16 cores each and 128GB of total memory, while each Summit node is a dual-socket IBM POWER9 with 22 cores each and 512GB of DDR4 from RAM, but only 42 cores are available because 2 of them are reserved for the operating system.
On Cori Haswell, we used \texttt{gcc-11.2} and the \texttt{O3} flag to compile C/C++ codes, while on Summit we used \texttt{gcc-9.2}.
On both machines, we used the default MPI implementation.

To evaluate our algorithm, we use three different species: Oryza sativa (O. sativa), Caenorhabditis elegans (C. elegans), and Homo sapiens (H. sapiens) whose characteristics are summarized in Table~\ref{tab:data}. 
The evaluation is divided into two categories: (a) runtime and scalability of the ELBA pipeline including our novel contig generation algorithm on two machines, and (b) runtime compared to state-of-the-art shared memory software.
For low error rate sequences (O. sativa and C. elegans), we also report assembly quality.

The state-of-the-art software we consider are Hifiasm~\cite{cheng2021haplotype} and HiCanu~\cite{nurk2020hicanu}, because of their speed and popularity, respectively, and both are written for shared-memory parallelism. 
HiCanu can optionally run on grid computing, but is not implemented for distributed memory parallelism.
It is worth noting that Hifiasm and HiCanu include additional polishing stages that make their overall assembly quality higher than ELBA's.
For the H. sapiens dataset, we consider Miniasm~\cite{li2016minimap} and Canu~\cite{koren2017canu}, as this dataset has a much higher error rate that is not suitable for Hifiasm and HiCanu.
Our goal is to show the competitiveness and potential of ELBA in terms of assembly quality, demonstrating in particular clear advantages in terms of runtime.
ELBA was run with the k-mer length parameter $k = 31$ and the x-drop threshold $x = 15$ for the low error rate data and with $k = 17$ and $x = 7$ for H. sapiens in Table~\ref{tab:data}.
Hifiasm and HiCanu were run with their default setting.

To show performance, we run ELBA on both Cori Haswell and Summit (except for H. sapiens, where we only use Summit), while for comparison with the state of the art, we only use Cori Haswell, since Hifiasm and HiCanu use SSE and AVX2 intrinsics, which are not supported on the IBM POWER9 processor on Summit.
Hifiasm and HiCanu were developed for shared memory, so we only give runtimes for a single Cori Haswell node using multi-threading.
The lack of support for AVX2 intrinsics is also the reason why ELBA's alignment is slower on Summit than on Cori.
% \Giulia{
For this reason, we want to emphasize that the goal of using two machines is to show performance and scaling on different systems, not to directly compare the two machines, since ELBA is optimized for a general HPC system, i.e., our code is general and no architecture-specific optimizations have been made.
%}

To assess the quality of the contig set of ELBA, Hifiasm, and HiCanu, we use QUAST~\cite{gurevich2013quast} and report the following metrics:
Completeness, longest contig size, number of contigs, and misassembled contigs.
Completeness measures the percentage of the reference genome to which at least one contig has been aligned. 
This is calculated by counting the number of nucleotides aligned to the reference genome and dividing by the total length. 
The number of misassembled contigs is defined as the number of contigs that contain incorrect assemblies, e.g., a contig consisting of sequences originating from different regions of the reference genome.
%These can be inversions (one region of the contig aligns to the reverse complement of the reference while flanking another region that aligns in the forward direction), translocations in which flanking regions align to different chromosomes, and relocations in which flanking regions align with a gap of more than 1K nucleotides.
%On Oryza Sativa, we used 

%%%\input{body/results}

\vspace{-.5em}
\section{Results}

In this section, we evaluate the performance and quality of the overall ELBA pipeline and our novel contig generation algorithm, both individually and compared to the state of the art.

\subsection{Performance and Scalability}

\begin{figure}
    \centering
    \includegraphics[width=\columnwidth]{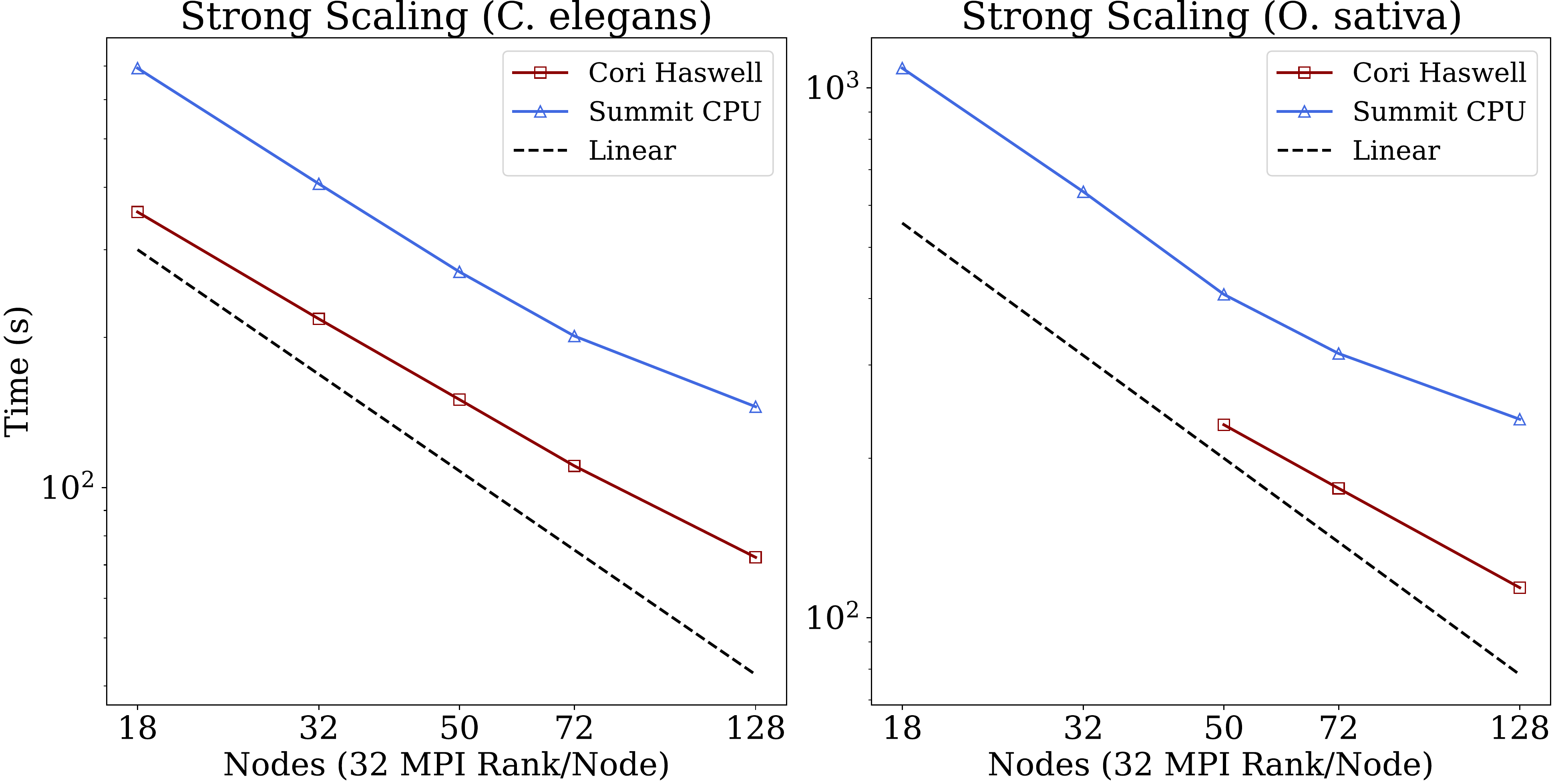}
    \caption{ELBA strong scaling on Cori Haswell and Summit CPU using 32 MPI rank/node on C. elegans (left) and on O. sativa (right).}
    \label{fig:ss-ce-os}
    \vspace{-.5em}
\end{figure}

\begin{figure}
    \centering
    \includegraphics[width=\columnwidth]{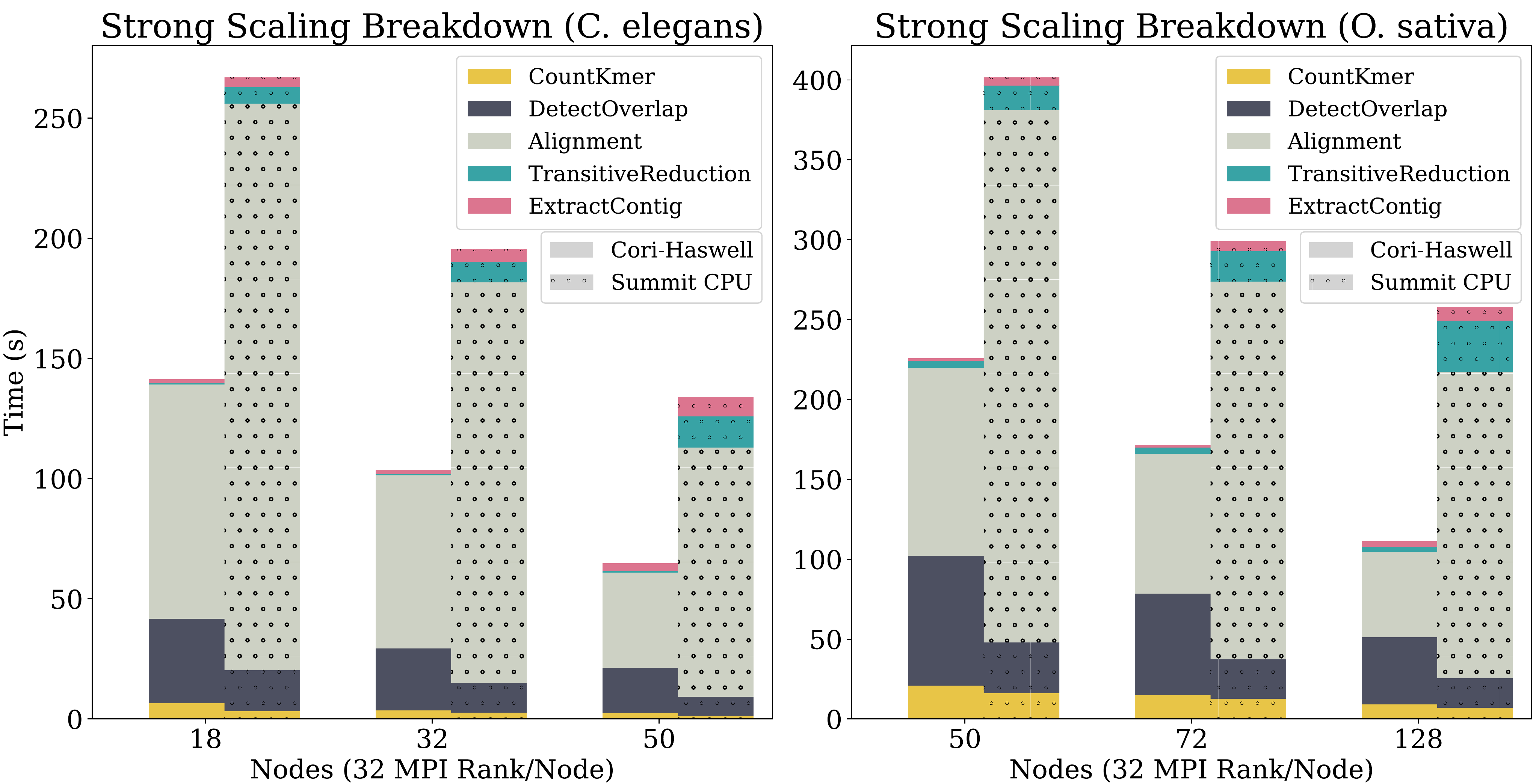}
    \caption{ELBA runtime breakdown of the main stages of the pipeline on Cori Haswell and Summit for C. elegans on the left and for O. sativa on the right.}
    \label{fig:rice-breakdown}
    \vspace{-1em}
\end{figure}

Figure~\ref{fig:ss-ce-os} illustrates the strong scaling of the entire ELBA pipeline for C. elegans on the left and for O. sativa on the right.
The C. elegans dataset was run on $P=$ \{18, 32, 50, 72, 128\} nodes using 32 MPI ranks/nodes on both machines, while the O. sativa was run on $P=$ \{18, 32, 50, 72, 128\} nodes using 32 MPI ranks/nodes on Summit CPU and on $P=$ \{50, 72, 128\} nodes Cori Haswell because the algorithm for $P=$ \{18, 32\} ran out of memory since Cori Haswell has a smaller memory per node than Summit.

\begin{figure}
    \centering
    \includegraphics[width=\columnwidth]{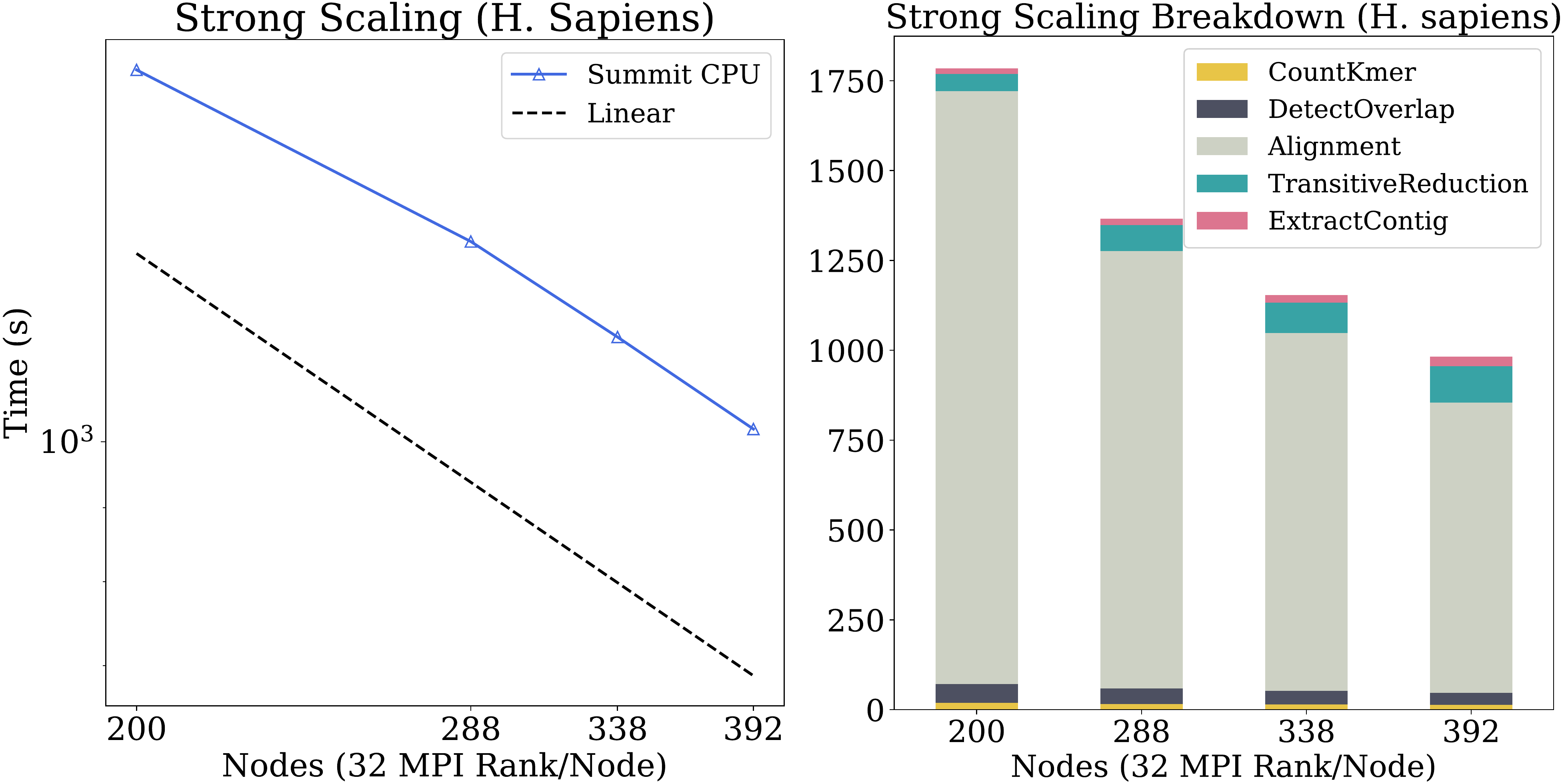}
    \caption{ELBA strong scaling (on the left) and runtime breakdown (on the right) of the main stages of the pipeline on Summit for H. sapiens.}
    \label{fig:human-breakdown}
    \vspace{-1em}
\end{figure}

% \Giulia{
ELBA achieves a parallel efficiency of 75\% on Cori Haswell and 69\% Summit CPU for C. elegans, while it achieves a parallel efficiency of 80\% and 64\% for O. sativa. The parallel efficiency of O. sativa between 72 and 128 nodes on Summit is 83\%, which is similar to the parallel efficiency on Cori between 50 and 128 nodes.
For the H. sapiens dataset, the parallel efficiency on Summit between 200 and 392 nodes is close to 90\%, as shown in Figure~\ref{fig:human-breakdown} on the left.
%}.
% \Giulia{On Summit, the alignment at the time of submission showed performance variability with parallel efficiency for O. sativa of 98-105\% for 18-72 nodes (N) and 69\% for 72-128N. Since submission, we performed additional runs that lowered variability: parallel efficiency at 72-128N of Summit is 83\%, which is similar to parallel efficiency on Cori at 50-128N. We will update the results with the average of all runs. New results for a Human dataset on 200-392N show scalability beyond 128N with parallel efficiency close to 90\% on Summit.}
These results show a good scaling behavior of the whole ELBA pipeline with a large input on two different architectures. 
%In Section~\ref{sec:discussion}, we will discuss the consistent discrepancy between Cori Haswell and Summit CPU in terms of scaling.

% Parallel Efficiency C. elegans Cori Haswell 74.63%
% Parallel Efficiency C. elegans Summit CPU 68.71%

% Parallel Efficiency O. sativa Cori Haswell 79.27%
% Parallel Efficiency O. sativa Summit CPU 59.49%

% Parallel Efficiency Yeast Cori Haswell 82.00%
% Parallel Efficiency Yeast Summit CPU 64.84%

% Parallel Efficiency H. sapiens Cori Haswell 89.72%
% Parallel Efficiency H. sapiens Summit CPU 81.61%

% \Giulia{
Figure~\ref{fig:rice-breakdown} shows the runtime breakdown of ELBA on the two machines for C. elegans on the left and for O. sativa on the right, while Figure~\ref{fig:human-breakdown} on the right shows the runtime breakdown of ELBA on Summit for H. sapiens.
%}
To highlight the impact of the major stages, we omit I/O and other minor computation from the breakdown. 
The overall impact of the omitted computation is negligible.

The legend is arranged in reverse order with respect to the layers of the bar, i.e. the first entry of the legend at the top is associated with the first layer in the stacked bar from the bottom.
\texttt{CountKmer} corresponds to the k-mer counting step, then follows the step \texttt{DetectOverlap}, which represents the time to create and compute the candidate overlap matrix $\mC$, i.e., the first step shown in Figure~\ref{fig:pipeline}.
Then comes the \texttt{Alignment} step, i.e., the time required to perform pairwise alignment on each nonzero in the candidate overlap matrix $\mC$, followed by \texttt{TrReduction}, i.e., the transitive reduction time and the second step shown in Figure~\ref{fig:pipeline}.
Finally, \texttt{ExtractContig} is the time we spend extracting the contig set from the sparse matrix $\mS$ and is shown as the third step in Figure~\ref{fig:pipeline}. 
The \texttt{ExtractContig} step is the core contribution of this work, but its implementation is an essential part of making the entire ELBA pipeline work and for this reason it is key to showing the scalability of the pipeline.

% \Giulia{
Figure~\ref{fig:rice-breakdown} shows the breakdown of ELBA performance for the main phases of computation for C. elegans and O. sativa using $P{=}\{50,72,128\}$ nodes on each machine while Figure~\ref{fig:human-breakdown} shows the breakdown of ELBA performance breakdown for H. sapiens using $P{=}\{200,288,338,392\}$ Summit nodes.
%}
ELBA is faster overall on Cori Haswell than on Summit CPU.
% \Giulia{
The \texttt{CountKmer}, \texttt{DetectOverlap}, and \texttt{Alignment} phases show nearly linear scaling on both machines, with the exception of \texttt{CountKmer} for H. sapiens on Summit, which scales only sublinearly.
%}
The relative contribution of pairwise alignment to the overall runtime increases on Summit CPU compared to Cori Haswell and largely contributes to the discrepancy in runtime because the alignment library used in ELBA is not optimized for the IBM processor.
The \texttt{TrReduction} and \texttt{ExtractContig} phases are also significantly faster on Cori Haswell than on Summit CPU. 
On Summit, the computation for these two phases does not scale and consumes a higher percentage of runtime. This is because the amount of work is smaller in these two phases and the algorithms are latency-bound. Further, Summit's network has lower performance compared to Cori Haswell's.
Summit CPU has lower network bandwidth per core, and we also only use 32 of the 42 available cores on Summit to make the comparison with Cori fair, which does not saturate Summit's bandwidth.

In each dataset, 65-85\% of the runtime of contig generation on both machines is taken by the induced subgraph function described in Section~\ref{fig:induced}, which mainly involves communication.
The \texttt{ExtractContig} never requires more than 5\% of the computation for each species and each machine, demonstrating the efficiency of our contig generation algorithm.

ELBA's contig generation focuses on localizing the graph traversal problem so that the \nobreak sequences that make up a contig are organized into localized matrices on each processor.
The working set at this stage is smaller than at the beginning of the computation, so the result is a fast and efficient, but latency-bound algorithm.
%and the algorithm focuses on maximizing local computation

% \Giulia{Contig gen only doesn't scale as we are communication bound, we should still have some of those numbers and explain why that is the case, i.e. our algorithm doesn't scale because is good, not because is bad.}

\vspace{-.5em}
\subsection{Comparison with the State-of-the-Art}\label{sec:soa}

% \Giulia{
To demonstrate the competitiveness and advantages of ELBA over state-of-the-art long read assembly software, we compare the runtime and scaling of ELBA with Hifiasm~\cite{cheng2021haplotype} and HiCanu~\cite{nurk2020hicanu} for O. sativa and C. elegans and with Mifiasm~\cite{li2016minimap} and Canu~\cite{koren2017canu} for H. sapiens.
%}
The competing software are designed for shared memory parallelism.
Therefore, we ran them on a single Cori Haswell node with multithreading, while we ran ELBA on $P= \{18, 50, 128\}$ with 32 MPI ranks/nodes.
Due to the hard-to-separate differences between these software, we decided to make a comparison based on total runtime. 
Hifiasm, HiCanu, Mifiasm, Canu, and ELBA perform very different computation, but they ultimately aim to solve the same problem.
% Besides, the total runtime is also what an end user would be most interested in. 
% Moreover, f
For the sake of completeness, we also compare the quality of assembly.
It is worth noting that both Hifiasm and HiCanu perform a polishing phase.
This can lead to a slight disadvantage in runtime, but ensures a better assembly quality.

Table~\ref{tab:speedup} summarizes the runtime performance of Hifiasm and HiCanu, in the rightmost column the speedup of ELBA over those software for 
$P= \{18, 128\}$ and $P= \{50, 128\}$ for C. elegans and O. sativa, respectively.
ELBA is up to $15\times$ faster than Hifiasm and up to $58\times$ faster than HiCanu for the C. elegans dataset, while the speedup for O. sativa, which is a larger genome, reaches $36\times$ over Hifiasm and up to $159\times$ over HiCanu.
% \Giulia{
On Cori, runtimes for Hifiasm and HiCanu for O. sativa are 17 and 64 minutes, respectively, while ELBA takes less than 2 minutes on 128 nodes. 
For C. elegans, runtimes of Hifiasm and HiCanu are approximately 1 hour and 5 hours, while ELBA requires 1 minute on 128 nodes. 
For H. sapiens, ELBA on 392 nodes on Summit takes 17 minutes, while Mifiasm on Cori Haswell takes 1 hour and 30 minutes with 32 threads and Canu ran out of time after more than 64 hours of runtime with 32 threads on Cori Haswell.
However, these times are from different architectures and are not directly comparable, so we do not give a speedup for H. sapiens. 
The results from O. sativa and C. elegans show that performance on Summit was worse than on Cori Haswell due to the less powerful processor on Summit.
%}

Finally, we compare the quality of the assemblies of ELBA with that of Hifiasm and HiCanu. 
Table~\ref{tab:quast1} summarizes metrics obtained from QUAST~\cite{gurevich2013quast}. 
Specifically, we show the completeness, i.e., the proportion of the reference genome that was covered by at least one contig (the higher, the better), then the length of the longest contig (the higher, the better), the size of the contig set (the lower, the better if associated with high completeness), and the number of misassemblies (the lower, the better).
In both species, ELBA has competitive genome completeness, especially in C. elegans, where it is higher than both Hifiasm and HiCanu, and misassemblies.
In ELBA, the contigs are significantly shorter than in the two competing software.
This is understandable, since ELBA does not currently perform a polishing step, which is reserved as future work.

%\input{tables/table5-soa}

% \usepackage{booktabs}
\begin{table}[t]
\caption{ELBA's speedup over state-of-the-art software. Hifiasm and HiCanu are designed for shared memory parallelism and were run on a single Cori node with multithreading.}
\centering
\resizebox{\columnwidth}{!}{
\begin{tabular}{|l|l|c|c|c|c|}
\hline
\textbf{Tool} & \textbf{Organism} & \textbf{Runtime (s)}  & \textbf{Nodes (32 MPI Rank/Node)} &\textbf{ELBA Speed-Up}  \\ 
\hline
\hline
Hifiasm & C. elegans & \hspace{0.5em}1,015.2 & 18--128 & \hspace{1em}3--15$\times$ \\
\rowcolor{Gray}
HiCanu &  C. elegans & \hspace{0.5em}3,819.0 & 18--128 & \hspace{0.5em}11--58$\times$\\ [-11pt]
& & && \\
\hline
\hline
Hifiasm & O. sativa & \hspace{0.5em}4,131.9 & 50--128 & \hspace{0.5em}18--36$\times$ \\
\rowcolor{Gray}
HiCanu & O. sativa & 18,131.0 &  50--128 &  78--159$\times$ \\[-11pt]
& & && \\
\hline
\end{tabular}}
\label{tab:speedup}
\vspace{-1em}
\end{table}

%\input{tables/table3-assembly}

% \usepackage{booktabs}
\begin{table}[t]
\caption{Comparison of assembler quality for O. sativa (top) and C. elegans (bottom). Hifiasm and HiCanu implement additional polishing stages to improve their metrics.}
\centering
\resizebox{\columnwidth}{!}{
\begin{tabular}{|l|c|c|c|c|c|}
\hline
\textbf{Tool} & \textbf{Completeness (\%)} & \textbf{Longest Contig (Mb)} & \textbf{Contigs} & \textbf{Misassembled Contigs}  \\ 
\hline
\hline
ELBA & 37.09 & \hspace{0.5em}0.172 & 6411 & 2 \\
\rowcolor{Gray}
Hifiasm & 26.94 & \hspace{0.5em}7.083 & 1661 & 1 \\
HiCanu & 25.94 & 37.523 & \hspace{0.5em}168 & 2 \\

\hline
\hline
ELBA & 98.93 & \hspace{0.5em}0.313 & 4287 & 5 \\
\rowcolor{Gray}
Hifiasm & 99.96 & \hspace{0.5em}6.438 &  \hspace{0.5em}133 & 0 \\
HiCanu & 99.90 & 18.332 & \hspace{1em}32 & 2 \\
\hline
\end{tabular}}
\label{tab:quast1}
\vspace{-1em}
\end{table}

\vspace{-.5em}
\section{Conclusions}

Recent advances in sequencing technologies have increased the need for high-performance approaches, as we can now generate more data at lower cost, which in turn requires higher computational resources to reconstruct high-quality assemblies in a timely manner.
In this paper, we presented the contig generation phase of the distributed-memory long-read assembler ELBA. 
%\Aydin{I changed this preceding sentence because you don't want to sell this as the ultimate ELBA paper}. the first implementation of the distributed memory OLC assembly paradigm \Aydin{I don't think this is correct as OLC paradigm was around before Illumina; I actually think you should just delete that part of sentence coming before this commen}. 
%ELBA, which is built on diBELLA 2D, is a distributed-memory long-read assembler that uses the OLC paradigm.

Contig generation is critical to assembly functionality, as it enables the construction of longer genomic sequences that represent a physical map of a region of a chromosome.
ELBA has achieved speed increases of up to $36\times$ and $159\times$ compared to two state-of-the-art software for the O. sativa dataset, opening the door for high-performance genome assembly.

ELBA and its contig generation step rely on distributed sparse matrices to first determine the contig set starting from a string graph. 
Then, using a greedy multiway number partitioning algorithm, it determines how the rows and columns of the sparse matrix representing DNA sequences are redistributed between processes so that sequences belonging to the same contig are stored locally on the same processor.
ELBA then uses such partitioning and the sparse matrix abstraction to implement the induced subgraph function and redistribute the sequences among the processes.
Finally, ELBA's contig generation step computes a local assembly step, i.e., the actual concatenation of sequences into a contig, on each processor independently and in parallel. 

Future work includes developing a polishing or scaffolding phase to further improve the quality of ELBA assembly.
One possibility is to once again use the sparse matrix abstraction to find similarities within the contig set and obtain even longer sequences.
In addition, we plan to reduce the memory consumption of ELBA so that we can assemble large genomes at low concurrency, and optimize ELBA for running in a cloud environment as high-performance scientific computing in the cloud becomes more popular~\cite{guidi202110,reed2022reinventing,mesnard2019reproducible}.
Finally, we plan to use GPUs in several stages, such as alignment~\cite{zeni2020logan,awan2020adept} and k-mer counting~\cite{erbert2017gerbil,nisa2021distributed}, to improve our performance and take full advantage of today's heterogeneous systems. %,ahmed2019gasal2

% \vspace{-.5em}
\begin{acks}
This work is supported by the Advanced Scientific Computing Research program within the Office of Science of the DOE under contract number DE-AC02-05CH11231, and by the Exascale Computing Project (17-SC-20-SC), a collaborative effort of the DOE Office of Science and the NNSA. We used resources of the NERSC supported by the Office of Science of the DOE under Contract No. DEAC02-05CH11231 and of the OLCF supported by the Office of Science of the DOE under Contract No. DE-AC05-00OR22725.
\end{acks}

%\vspace{-.5em}
\balance{}
\bibliographystyle{ACM-Reference-Format}
\bibliography{references}

\end{document}